\documentclass[aps,pre,twocolumn,showpacs,reprint]{revtex4-1}

\usepackage{graphicx}% Include figure files
\usepackage{dcolumn}% Align table columns on decimal point
\usepackage{bm}% bold math
\usepackage{amssymb,amsfonts,amsmath}
\usepackage{graphicx,subfigure,amsmath,bbm}
\usepackage{color}
\usepackage{rotating}
\usepackage[sort&compress]{natbib}
\usepackage{tikz}

\begin{document}

%\preprint{APS/123-QED}
\title{Shear shock waves are observed in the brain}

\author{David Esp\'indola}
 \email{david.espindola@unc.edu}

\author{Stephen Lee} 
\author{Gianmarco Pinton}
 \email{gfp@unc.edu}

\affiliation{Joint Department of Biomedical Engineering, University of North Carolina at Chapel Hill and North Carolina State University. 109 Mason Farm Rd, CB 7575, Chapel Hill, NC 27514, USA}

\date{\today}

\begin{abstract}
The internal deformation of the brain is far more complex than the rigid motion of the skull. An ultrasound imaging technique that we have developed has a combination of penetration, frame-rate, and motion detection accuracy required to directly observe, for the first time, the formation and evolution of shear shock waves in the brain. Experiments at low impacts on the traumatic brain injury scale demonstrate that they are spontaneously generated and propagate within the porcine brain.  Compared to the initially smooth impact, the acceleration at the shock front is amplified up to a factor of 8.5. This highly localized increase in acceleration suggests that shear shock waves are a fundamental mechanism for traumatic injuries in soft tissue.
\end{abstract}

\pacs{Valid PACS appear here}

\maketitle
Traumatic brain injuries (TBI's) are a major source of death and disability worldwide. Falls and motor vehicle related accidents are the largest contributors. Current biomechanical predictive criteria for TBI are based on measurements of skull motion such as linear and rotational acceleration~\cite{rimel1981disability,rowson2013brain}. Although the relationship between skull motion and injury has been extensively tested~\cite{rowson2013brain,Broglio2009,Crisco20112673,Daniel2012,Laksari20150331,Wright2012}, mechanisms relating the two have not been conclusively established~\cite{guskiewicz2007measurement}, due to the complexity of the deformation of the brain~\cite{bayly2005deformation,MossBrain,GorielyBrain} which behaves as a nonlinear viscoelastic medium. {\it In situ} measurements of the rapid transient motion of the whole brain during a traumatic event may establish a more accurate biomechanical description of injury.

Shear vibration experiments on small brain samples have shown that the stress-strain relationship behaves nonlinearly for amplitudes as low as $1\%$~\cite{darvish2001nonlinear,Donnelly1997}, which is well below the strain threshold for injury~\cite{Wright2012,margulies1992,Sullivan2015}. Therefore, even mild injuries typically occur within the nonlinear elastic regime. Furthermore, the brain's shear wave speed ($c_t\sim$2~m/s) is three orders of magnitude smaller than the compressional wave speed ($c_p\sim$1500~m/s) therefore the deformation from an impact is almost entirely in shear mode. 

Outside of ultrasound, current methods that measure brain motion cannot capture this nonlinear behavior of shear waves due to limitations in either, frame rate, penetration, or motion detection accuracy. For instance, magnetic resonance elastography (MRE) has been used to investigate shear wave propagation in the brain, typically at frame rates on the order of tens of images/second. Furthermore the temporal sampling of MRE is fundamentally limited to a few ms by the finite integration time over the spin relaxation~\cite{hamhaber2010vivo,johnson2013magnetic}. The harmonics generated by nonlinear shear wave propagation exceed these sampling capabilities. Optical methods have larger frame rates and they can measure motion in optically transparent materials~\cite{Razani2012, Song2013, Razani2014}. However, they are limited to shallow penetration depths in soft tissue ($\sim2$~mm)~\cite{Park28072015}. 

The only experimental corroboration of nonlinear shear waves and its characteristic harmonic signature was made over a decade ago in a homogeneous gelatin phantom~\cite{catheline2003observation}. These experimental methods were derived from ultrasound elastography~\cite{tanter2002ultrafast} and performed with a high frame-rate (3000~images/second) ultrasound scanner and a cross-correlation tracking algorithm that measured the inter-frame displacement. The high imaging rate was required to capture the rapid, transient, and broadband nature of the shocks. At these high frame-rates, the movement between successive acquisitions is small ($\sim 1~\mu$m) and highly accurate motion measurement is required. Although these gelatin phantoms were calibrated to have the same linear elastic properties as soft tissue (Young's modulus) they did not mimic the nonlinear and attenuating properties of brain, which are key components needed to characterize the shock wave physics.

Despite this experimental report of shear shock waves in a homogeneous gelatin, there have been no reports of shear shock waves in the human body, including the brain. This is due to the difficulty of observing nonlinear shear displacements at depth and at high frame-rates in the brain which include: 1) the more challenging imaging environment, arising from the heterogeneous acoustical properties of its biological structures, and 2) to the higher attenuation of the shear waves. To measure the shear shock wave propagation in {\it ex vivo} porcine brain presented in this letter, we address these two challenges. We have designed a focused high frame-rate ultrasound imaging sequence that has better penetration and is more accurate compared to previous plane-wave methods and it is described here for the first time. This data is then processed with an adaptive tracking algorithm that can accurately characterize the almost discontinuous velocity profile at the shock fronts (see appendix C and~\cite{pinton2014adaptive}). With these advancements, motion is detected at high frame-rates and with high accuracy for all points within an imaging field of view as deep as the brain. It is shown that shear shock waves are spontaneously generated in the brain, i.e. a smooth excitation at the brain surface develops into a shock as it propagates within the brain. We show that the strong gradients, at the shock front, dramatically increase the local acceleration, strain and strain-rate in soft tissue which suggests that this is a primary injury mechanism for a broad range of TBI's. 

\begin{figure}[h!]
\centering
\includegraphics[trim=0 0 0 0,clip,width=0.47\textwidth]{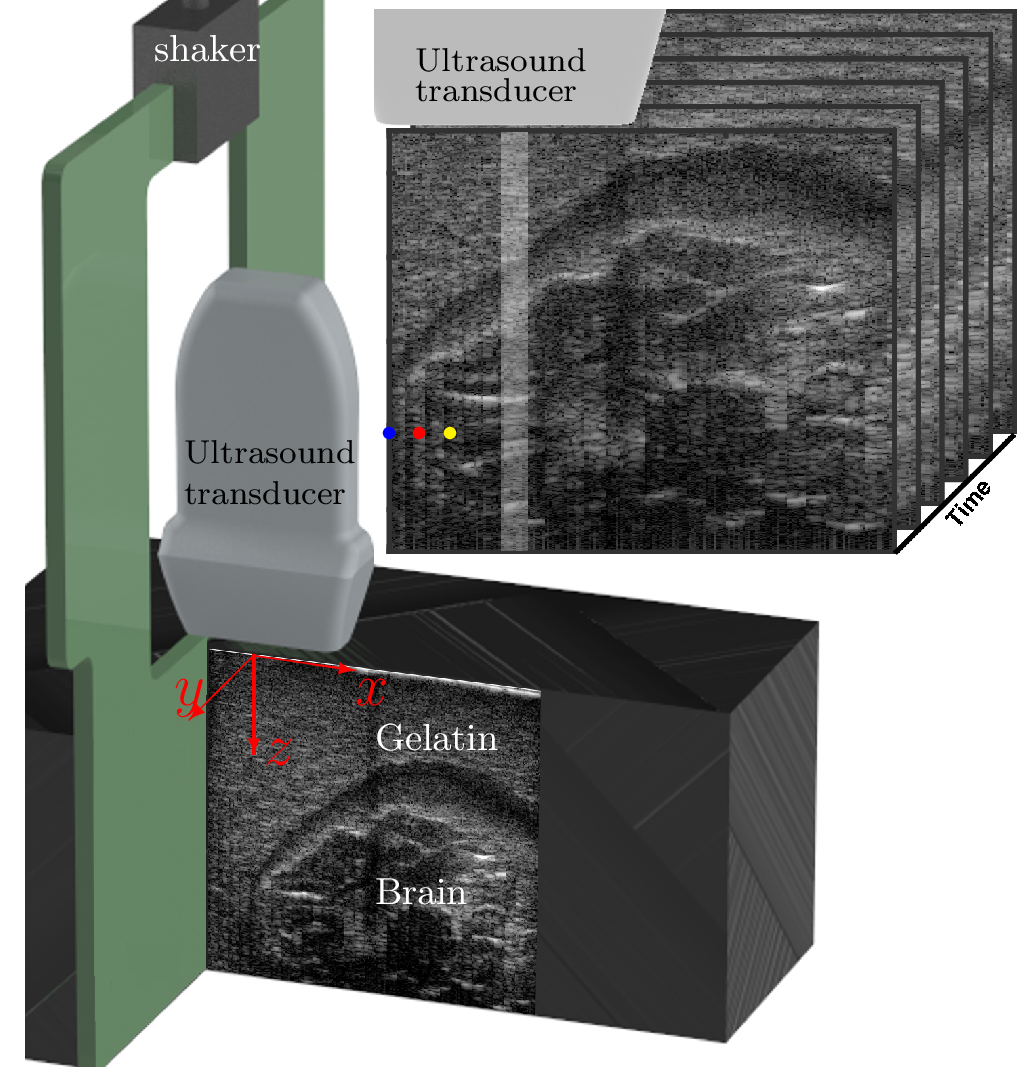}
\caption{(Color online). A plate is embedded in gelatin-brain to transmit shear waves with an electromechanical shaker. The shear wave produces displacement along depth ($z$) and it propagates along the $x$ axis. The Bmode sequence was acquired from 32 synchronized focused transmit-receive events. The focal depth was kept constant (60~mm) and each lateral focus position was shifted. The highlighted vertical stripe shows the region corresponding to a single transmit-receive event~[\citenum{sup}a].}
\label{fig:ExpSetup}
\end{figure}

The experimental design used to generate and measure shear shock waves in the brain is illustrated in Fig.~\ref{fig:ExpSetup}. Porcine brains (8 week old Yorkshire cross pigs), were extracted 1 hour post-mortem to minimize tissue degradation and were embedded in gelatin. Graphite powder (10-100~$\mu$m) was added to the gelatin with a concentration of 1~\% to generate acoustical contrast for the ultrasound images. The shear wave speed of the gelatin was calibrated by setting the concentration of 225 bloom gelatin to 5.0\% by weight. This yielded a shear wave speed equivalent to measurements performed in porcine brains ($2.14\pm0.06$~m/s~, see appendix A]). The speed of sound was calibrated by adding 5.0\%  isopropyl alcohol to obtain a speed of $1490$~m/s~\cite{duck1990physical}. Before the gelatin cured, a 6~mm thick acrylic plate was embedded near the surface of the brain, as shown in Fig.~\ref{fig:ExpSetup}. Then, the plate was connected to an electromechanical shaker and the gel was allowed to cool and cure in an ice water bath.

By matching the shear and compressional wave speeds between the brain and gelatin a high transmission coefficient was obtained at the gelatin-brain interface. This is a non-biofidelic interface since {\it in vivo} the impact energy has to be transmitted through layers of skull, meninges, cerebrospinal fluid, etc. The effect of the skull's shape is also ignored, although simulations indicated that the spherical nature of the skull facilitates the formation of shear shock waves deep in the brain by virtue of its natural focal geometry~\cite{giammarinaro2014numerical}. However, the complex boundary conditions at the skull-brain interface would generate wave-fields that are currently not solvable by either theoretical or numerical methods in the nonlinear regime. Instead, narrow band planar waves provide conditions for which theoretical results are known. Typically impacts that cause injury would not generate a
pure polarized shear wave. Furthermore they have a short duration and a broad power spectrum.Here we generated a 75~Hz wave train, which is within the range of frequencies measured during head impacts (10-300~Hz)~\cite{Laksari20150331,greenwald2008head}. It has 5 cycles and a -80~dB Chebyshev envelope. This spectrally narrow signal has good separability in the frequency domain which is necessary for the observation of the harmonic development and its comparison to theoretical predictions of shear shock wave development. Note that the physics of shock wave steepening and formation remains applicable to other loading conditions, such as impulse-loading. This pulse was sent to a power amplifier to generated amplitudes up to 450~m/s$^2$ as measured by a linear accelerometer attached to the plate. This setup generates linearly polarized shear plane waves with a full control of its frequency content. The shear wave induces a movement in the $z$-axis and it is propagated through the $x$-axis.

Shear wave propagation through the brain and gelatin was monitored with a research ultrasound scanner (Verasonics). A 5~MHz commercial diagnostic ultrasound imaging probe (ATL Philips L7-4) was positioned above the phantom, as shown in Fig.~\ref{fig:ExpSetup}, using a high precision (10~$\mu$m) robotic arm to monitor different sections of the brain. 

\begin{figure}[h!]
\centering
\includegraphics[trim= 0 0 0 0, clip,width=0.48\textwidth]{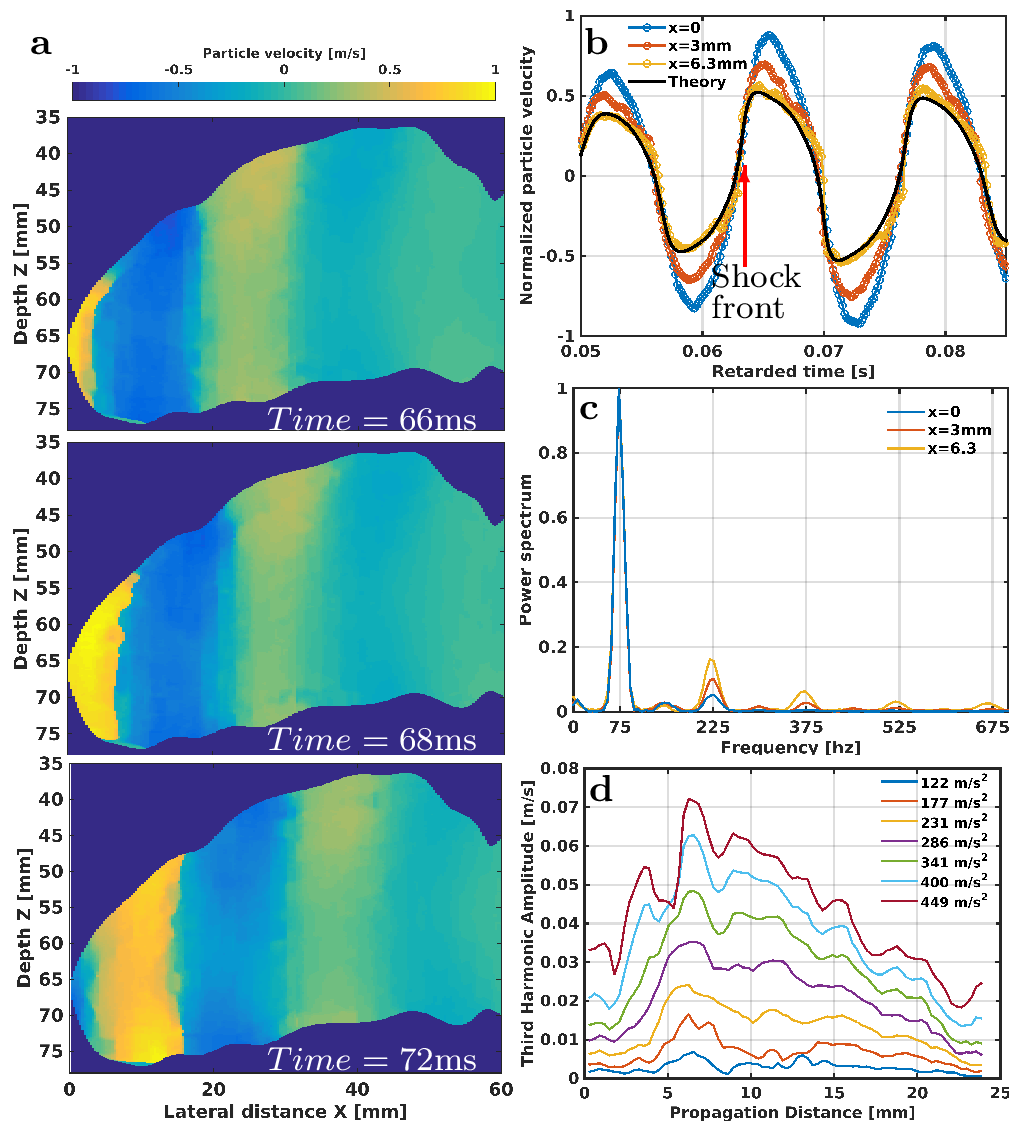}
\caption{(Color online) Snapshots of particle velocity generated by a planar shear wave propagating in the brain~[\citenum{sup}b] ({\bf a}). The initial condition was 449~m/s$^2$. The evolution of shear wave propagation in the time domain ({\bf b}) shows excellent agreement with theoretical predictions for a homogeneous medium (black line). The specific odd harmonic development is visible in the frequency domain ({\bf c}). The curves in {\bf b} and {\bf c} correspond to the motion at single points represented by circles in Fig.~\ref{fig:ExpSetup}. As the driving amplitude increases so does the amplitude at the third harmonic frequency ({\bf d}).}
\label{fig:velMap}
\end{figure}

Conventional planar and plane-wave compounding~\cite{montaldo2009flashangles} were sensitive to off-axis clutter and did not perform optimally for {\it ex vivo} brain tracking. Sources of ultrasound image degradation, such as reverberation~\cite{pinton2011sources} and bright off-axis targets which generated static and dynamic artifacts, decreased the image quality and degraded motion tracking performance. This motivated the development of a new sequence which consists of 16 transmit-receive events focused at different lateral positions, each one with an F-number of 4, and a focal depth of 6~cm. Each transmit-receive event generated 1200 frames of raw acoustical data. Then, a delay-and-sum beamforming algorithm generated, from every event, 8 A-lines of focused RF data (see the highlighted area in Fig.~\ref{fig:ExpSetup}). To maintain a high frame rate each transmit-receive event was synchronized to the shear wave generation. Furthermore, to increase the visualization field, the acoustical probe was mechanically displaced to a second position using the robotic arm (see appendix B). Thus, the images shown in this letter were generated from 32 synchronized transmit-receive sequences, with an effective frame rate of 6200 frames/second. The jitter error on the synchronization time was measure to be 380~ns which is equivalent to a 1/35000 of a period at 75~Hz. The experiment was performed in four different brains. For each brain, seven amplitudes were used, ranging between 122-449~m/s$^2$. Five independent realizations were obtained for each amplitude by translating the transducer along the $y$-axis yielding a total of 35 experiments for each brain. All the experiment were performed within 15 hours post-mortem at a temperature of 1$^{o}$C.

The ultrasound data was used to generate a high frame-rate B-mode movie that illustrates the brain anatomy and the movement induced by the shear deformation (Fig.~\ref{fig:ExpSetup})~[\citenum{sup}a]. We determined brain motion based on the beamformed RF data with a motion detection algorithm designed to track the large discontinuous velocities associated with shear shock waves~\cite{pinton2014adaptive}. These shear excitations result in inter-frame deformations that change the relative scatterer positions and reduce the correlation between frames. The tracking algorithm uses a quality weighted three dimensional median filter to iteratively optimize the correlation values and it was calibrated to preserve the high frequency displacements necessary to characterize the sharp shock front (see appendix C for validation using simulated ultrasound imaging~\cite{pinton2012three,Pinton2016smalldisplacement} and Rusanov solutions of cubically nonlinear shocks). %This tracking algorithm was validated and optimized by simulating ultrasound imaging  with a finite difference method~\cite{pinton2012three,Pinton2016smalldisplacement} and by simulating the nonlinear elastic motion of the brain with a Rusanov scheme that was designed to capture cubically nonlinear shock behavior~[\citenum{sup}d].

Three snapshots of the resulting high frame-rate movie of particle velocity~[\citenum{sup}b] are shown in Fig.~\ref{fig:velMap}{\bf a} for the case where the amplitude of the particle velocity at the plate was $v_0=1$m/s which produced a Mach number of $M=c_T/v_0=0.49$. A quasi-planar wave propagating from left to right can be clearly observed. Due to the high Mach number, the brain develops a nearly discontinuous velocity profile in Fig.~\ref{fig:velMap}{\bf b}. 

Describing the brain as a nonlinear elastic solid is theoretically complex due to high order tensors in the strain energy density~\cite{landau1960theory,Wochner2008}. By assuming a linear polarization the general description can be reduced to a simpler scalar representation~\cite{zabolotskaya2004modeling}, while still retaining the nonlinearities that produce complex waves physics~\cite{pinton2010nonlinear,whitham1974}. For a plane polarized shear wave the equation of motion, for the particle velocity $v(x,t)$, can be written as:
\begin{equation}
\frac{\partial v}{\partial x}-\frac{\beta}{c_t^3}v^2\frac{\partial v}{\partial \tau}=\delta\frac{\partial^2 v}{\partial \tau^2},
\label{eq:rusanovic}
\end{equation}
\noindent where $\tau=t-x/c_t$ is the retarded time respect to the physical time $t$, $x$ is the propagation distance and $\beta$ quantifies the nonlinearity. The term on the right hand side accounts for dissipation, which has a significant influence on the shock dynamics. Equation~\ref{eq:rusanovic} has similar form to the well known Burgers' equation~\cite{burgers1948mathematical}, except that the nonlinear term is cubic rather than quadratic. This equation is almost unstudied from the experimental point of view due to the difficulty of directly observing shear waves at depth in solids. As the wave propagates the peaks and troughs are distorted by the cubic nonlinearity which locally increases the shear wave speed as a function of amplitude. Therefore, the wave will eventually ``tip'' over forming shocks for both the positive and negative wave phases. The cubic nonlinearity has a recognizable impact on the spectral development that occurs with propagation because it produces only odd harmonics.

The experimental development of this nonlinearity and its very specific odd harmonic signature can be observed in the particle velocity (Fig.~\ref{fig:velMap}{\bf b}) and power spectrum (Fig.~\ref{fig:velMap}{\bf c}). {\it This is direct evidence of shear shock wave development in the brain.} The Rusanov scheme, along with previously measured frequency dependant attenuation, were used to fit the experimental data at $x=$6.5~mm to the equation~\ref{eq:rusanovic}. This allowed us to estimate $\beta\approx13\pm6$ producing an excellent agreement between the theory(black curve) and experiment~\cite{rusanov1970method,pinton2010nonlinear}. In the frequency domain, consistently with theoretical predictions, energy is transferred from the fundamental frequency to the odd harmonics. Furthermore, more energy is sent to the odd harmonics as the initial wave amplitude increases (Fig.~\ref{fig:velMap} {\bf d}). The third harmonic amplitude as a function of the propagation distance (Fig.~\ref{fig:velMap} {\bf d}) indicates that the peak nonlinearity occurs at a propagation distance between 6-7~mm that is independent of initial velocity. Beyond this distance, viscosity overcomes nonlinear harmonic generation. The close match between the theoretical prediction and observation indicates that nonlinearity is a first order parameter required to accurately describe shearing mechanisms in the brain.

\begin{figure}[t!]
\centering
\includegraphics[trim= 0 0 0 0, clip,width=0.48\textwidth]{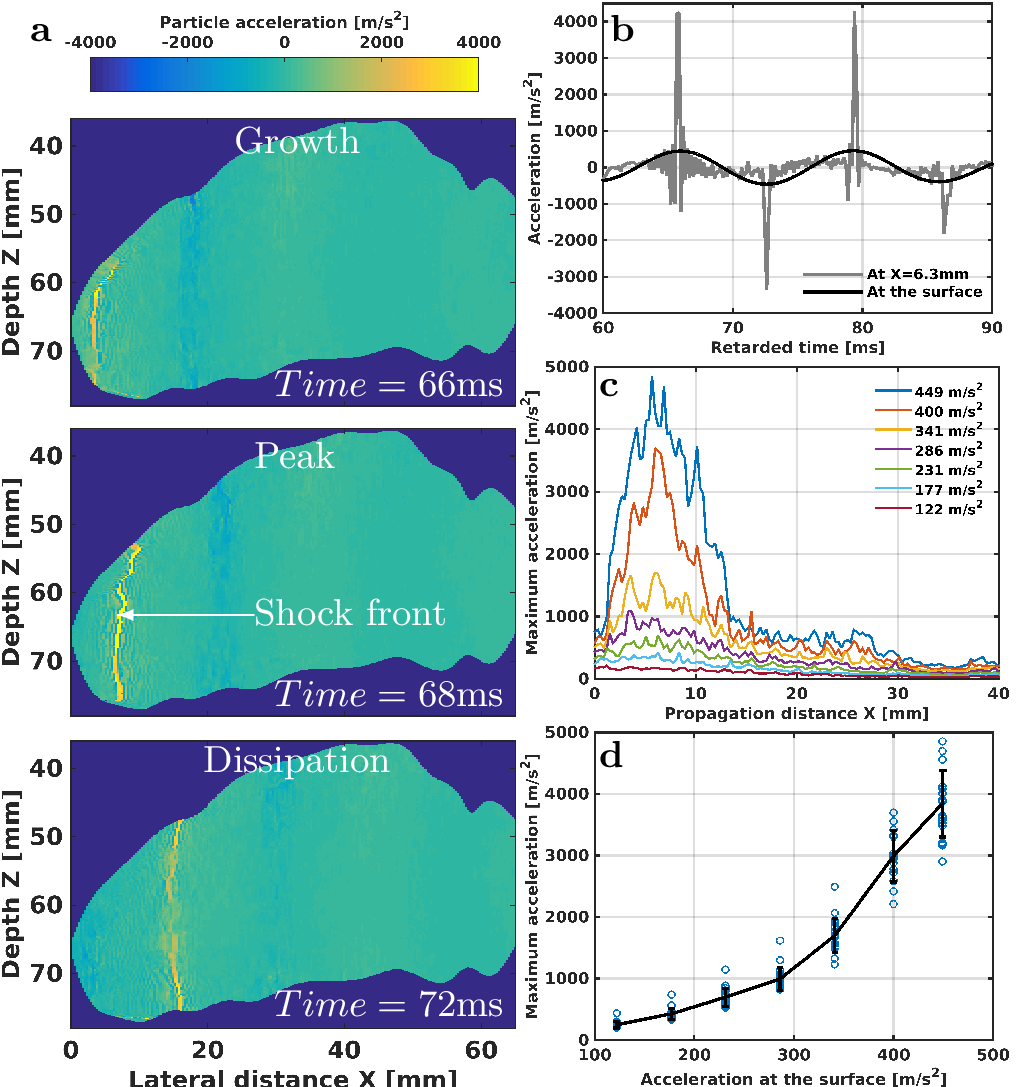}
\caption{(Color online). {\bf a} Snapshots of the particle acceleration showing growth, peak, and dissipation of the shock front. {\bf b} Time waveform of the acceleration signal at the surface of the brain and at 6.3mm for an initial amplitude of $449$m/s$^2$. {\bf c} The peak acceleration as a function of propagation distance. {\bf d} Maximum peak acceleration as a function of acceleration at the surface of the brain, the circles represent the measurements of the four brains and the five independent imaging slices. The solid line is the average curve and the error-bars shows the standard deviation.}
\label{fig:AccMap}
\end{figure}

Note that, in disagreement with Eq.~\ref{eq:rusanovic}, a significant but small portion of the fundamental energy is also transferred to even harmonics. This may be due to the brain structure, which is heterogeneous, and therefore violates the assumption of homogeneity. In fact it is surprising that the brain heterogeneity has so little influence on the harmonic development. We believe that this is a consequence of the shear wavelength, 28.5$\pm$0.8mm at 75~Hz, which is larger than the characteristic structures in the porcine brain such as vessels and sulci and which therefore leads to weak wave--brain structure interactions.

\begin{figure}[t!]
\centering
\includegraphics[trim= 0 0 0 0, clip,width=0.48\textwidth]{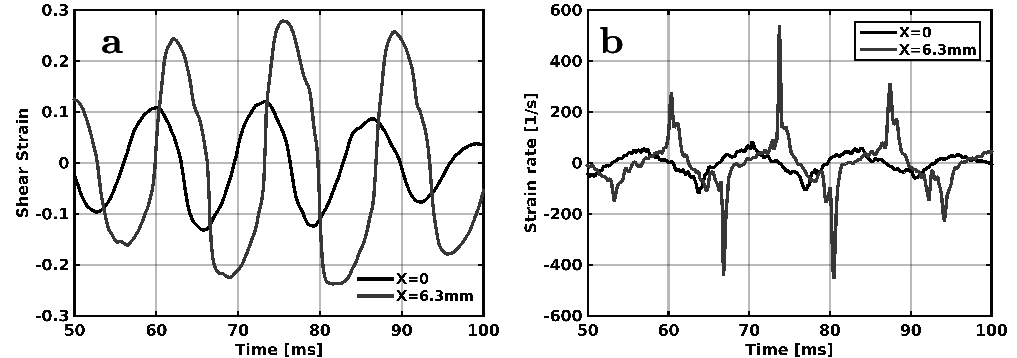}
\caption{Strain {\bf a} and strain-rate {\bf b} at the surface (black) and $x=6.3$mm (gray). The initial condition was $449$m/s$^2$.}
\label{fig:strain}
\end{figure}

The particle acceleration, which is more relevant to brain injury than particle velocity~\cite{rowson2013brain}, was calculated with a broadband derivative in frequency~\cite{oppenheim1983signals}. Numerical differentiation is noise sensitive, especially at the sharp shock profiles (see Fig.~\ref{fig:velMap}b) and due to false peak errors from of the tracking algorithm~\cite{pinton2014adaptive}. A number of validation steps were therefore performed to determine the error in the derivative calculations (see appendix D). They are shown to be in a good agreement with respect to numerical simulations. 

The resulting particle acceleration has a large amplitude that is finely concentrated at the shock front as it propagates into the brain (Fig.~\ref{fig:AccMap}{\bf a} and~\citenum{sup}c]). The shear wave grows, peaks, then dissipates within the first few centimeters of propagation. As a shear wave with an initial acceleration of 449~m/s$^2$ propagates into the brain  the steepening of the shock front amplifies the velocity gradient until it reaches 4200~m/s$^2$ at its peak, at $x=6.3$~mm of propagation  (Fig.~\ref{fig:AccMap}{\bf b}). Thus the acceleration in the brain is 9.3 times greater than the acceleration at the surface of the brain. The peak acceleration into the brain consistently occurs at $x=6.3$mm for the range of initial conditions considered here (Fig.~\ref{fig:AccMap}{\bf c}). Furthermore the maximum acceleration within the brain increases rapidly as the surface excitation increases, i.e. it is strongly nonlinear (Fig.~\ref{fig:AccMap}{\bf d}). At the maximum surface excitation the average amplification factor is 8.5.

In addition to acceleration, shear strain and strain rate are often used to predict brain injury~\cite{Wright2012,Sullivan2015}. Both are shown in Fig.~\ref{fig:strain}. The shear strain has an amplitude of 0.12 and 0.28 at the surface and at 6.3mm, i.e. an amplification factor of 2.3. The nonlinear distortion of the strain with propagation, which is similar to what was observed for the velocity, generates a dramatic increase in the strain-rate at the shock front (cf. Fig.~\ref{fig:strain}{\bf b} and Fig.~\ref{fig:velMap}{\bf b}). This generates an amplification factor of 6.4 with respect to the brain surface. These shear strain and strain-rate amplitudes are above the thresholds for injury estimated for TBI (0.07 of strain and 40s$^{-1}$ for strain-rate)~\cite{Sullivan2015}. 

In a real life impact, the skull, reflective fluid-solid interfaces and non-polarized initial conditions will generate more complex and disorganized shear waves than a plane wave. In this highly complex scenario, we think that at peak locations a shock front will be locally generated, magnifying the local acceleration, shear strain and strain-rate experienced by neuronal tissue. This etiology is consistent with diffuse axonal injury, one of the most common and devastating forms of TBI, which consists of numerous small (1-15mm) lesions that are distributed deep in the brain, far from the area of impact.

In conclusion, by directly measuring brain motion with high frame-rate ultrasound techniques, we have demonstrated that the combination of nonlinearity and attenuation is sufficient to develop shear shock waves over short propagation lengths starting from a smooth, low impact, shear wave in the {\it ex vivo} porcine brain. The physical mechanism from which this shear shock wave is generated is consistent with theoretical predictions of cubic nonlinearities that have a specific and readily identifiable odd harmonic signature. The imaging techniques and shock wave tracking methods developed here are not restricted to the brain, and can be applied to soft tissue anywhere in the body that is accessible to conventional ultrasound imaging.

\begin{acknowledgments}
The authors would like to acknowledge funding from the National Institutes of Health (R01 06052014). They would also like to thank Gregg Trahey and Francisco Melo for their comments on the manuscript.
\end{acknowledgments}

\section*{APPENDIX A: Mechanical characterization of the brain}

\begin{figure}
 \centering
\includegraphics[trim= 0 0 0 0, clip,width=0.48\textwidth]{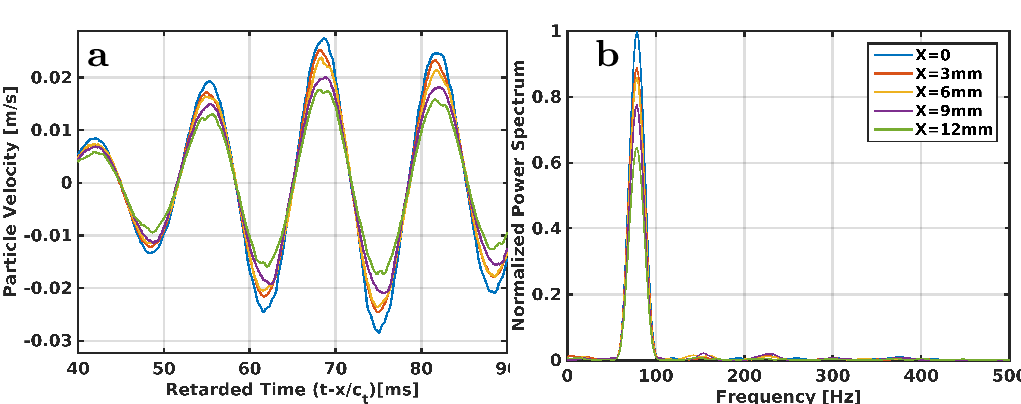}
\caption{Particle velocity ({\bf a}) and power spectrum ({\bf b}) generated by a linear shear wave traveling through the brain for different propagation distances.}
\label{fig:lin}
\end{figure}

Nonlinear propagation transfers energy from a signal's fundamental frequency to higher frequencies via harmonic generation. To avoid the complex frequency dependent interaction between nonlinearity, attenuation, and velocity, measurements of the mechanical properties of the brain had to be performed in the linear regime. To characterize these linear mechanical parameters of the brain, a low amplitude $2$m/s$^2$, $75$Hz excitation was used to generate a shear wave. The particle velocity for different propagation distances in the brain, $x = 0$, $3$, $6$, $9$, $12$mm (Fig.~\ref{fig:lin}) show that as the shear wave propagates in the brain, there is a decrease in amplitude but no measurable harmonic generation. This confirms that subsequent measurements of attenuation and dispersion were performed in the linear regime, and are independent of the effects of nonlinearity.

To measure the linear mechanical properties of the brain, the amplitude ($A$) and the phase ($\phi$), as a function of distance of propagation ($x$), were extracted from the Fourier transform of the experimentally measured particle velocity for the fundamental frequency. The attenuation, $\alpha$, was obtained by fitting the expression $A(x) = e^{-\alpha x}$ to the measured amplitude. The average characteristic attenuation length was found to be $L_A=1/\alpha=25\pm2$mm. The shear speed, $c_t$, was determined by fitting the experimentally measured phase to a linear phase shift, given by $\phi(x)=x\omega/c_t+\phi_0$, where $\omega$ is the angular frequency and $\phi_0$ is an arbitrary reference phase. The average shear speed was found to be $2.14\pm0.06$m/s. Here the error represents the standard deviation across the four brain samples.

\section*{APPENDIX B: Imaging sequences and validation}

\begin{figure}[h!]
\centering
\includegraphics[trim= 0 0 0 0, clip,width=0.48\textwidth]{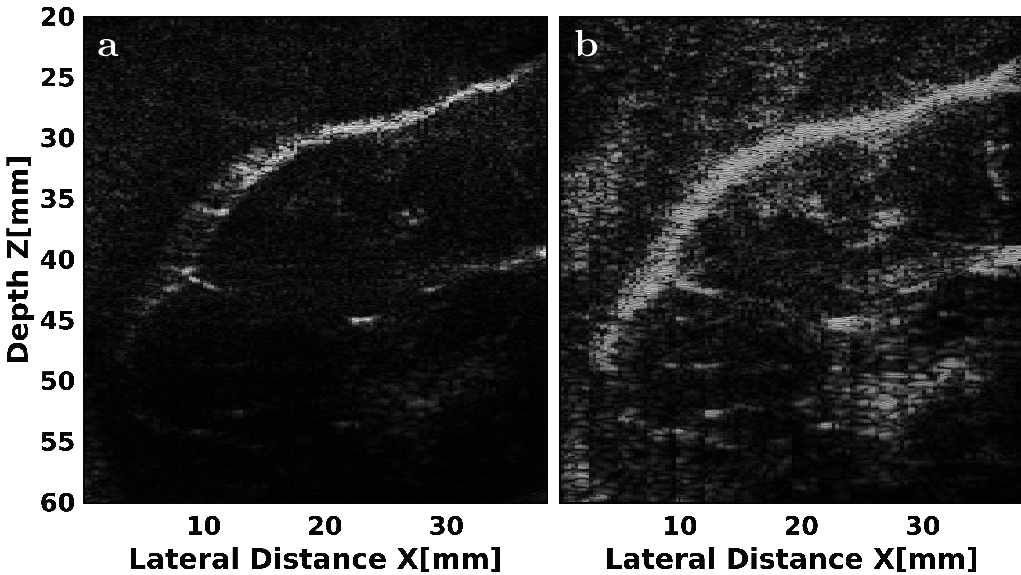}
\caption{Comparison of the B-mode images obtained from a conventional plane-wave compounding ({\bf a}) and a flash focus sequences ({\bf b}).}
\label{fig:Bmode}
\end{figure}

\begin{figure}[h!]
\centering
\includegraphics[trim= 0 0 0 0, clip,width=0.48\textwidth]{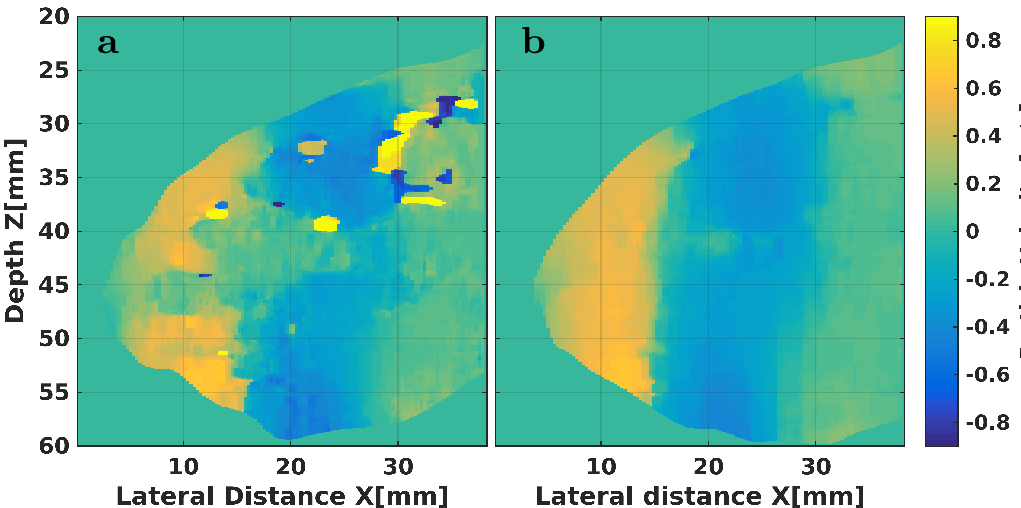}
\caption{Comparison of the paticle velocity measurement obtained from a conventional plane-wave compounding ({\bf a}) and a flash focus sequences ({\bf b}). Note that the color scale for both images is the same, and few regions in A are out of range.}
\label{fig:vel}
\end{figure}

To validate the proposed shock tracking ultrasound imaging sequence, referred to as flash focus, two experiments were performed in the brain with a driving amplitude of $449$m/s$^2$. The first imaging sequence consisted of a flash focus sequence at one transducer position. The second sequence consisted of the gold-standard conventional plane-wave compounding scheme \cite{montaldo2009flashangles}, each plane wave had a different angle of propagation in the $x-z$ plane. To be comparable to the flash focus acquisition time we used $15$ angles and one transmit-receive event per angle. These angles were chosen to be equally spaced between $\pm18^{\circ}$ with respect to the $z$-axis. Each angle was synchronized with the shear wave emission. After the multiple angles were acquired, the data was beamformed to generate $1200$ frames of RF data at $6200$ frames per second.

\begin{figure}
 
\centering
\includegraphics[trim= 0 0 0 0, clip,width=0.48\textwidth]{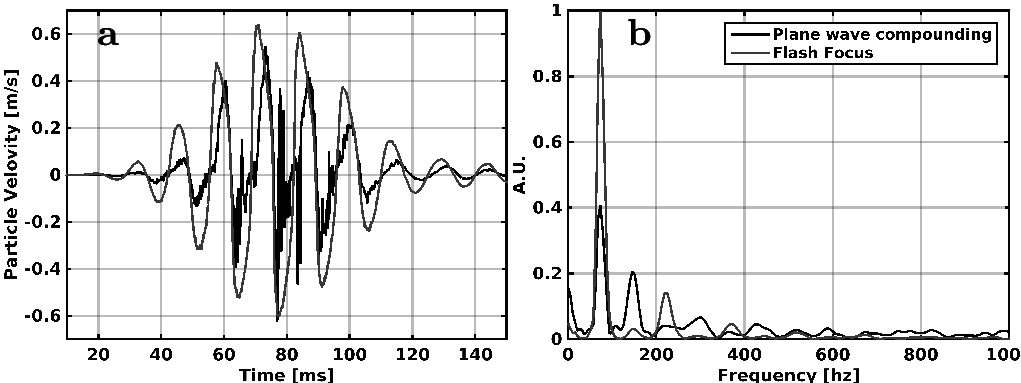}
\caption{Plot {\bf a} shows the time domain waveform and its corresponding frequency spectrum {\bf b} for the plane-wave compounding and flash focus sequences at $45$mm of depth and at a propagation distance of $X=10$mm.}
\label{fig:Sig}
\end{figure}

\begin{figure}
 \centering
\includegraphics[trim= 0 0 0 0, clip,width=0.48\textwidth]{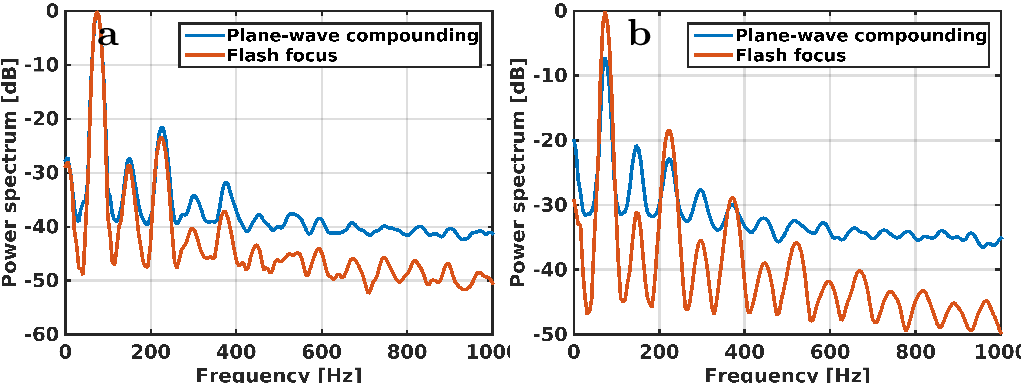}
\caption{Average power spectrum measure in the gelatin({\bf a}) and in the brain brain ({\bf b}) with the plane wave compounding and with the flash focus sequences.}
\label{fig:fft}
\end{figure}

Fig \ref{fig:Bmode} shows the B-mode image corresponding to the conventional plane-wave compounding ({\bf a}) and the  flash focus sequences ({\bf b}). The interframe displacement was calculated from the beamformed RF data with the same adaptive tracking algorithm for both cases and the resulting particle velocity is shown in Fig. \ref{fig:vel} {\bf a}, for plane wave compounding, and {\bf b}, for flash focus. It is qualitatively apparent from this figure that the flash focus sequence provides a more uniform estimate of motion that is consistent with a planar shear wave. The time plot, shown in Fig. \ref{fig:Sig} {\bf a}, is smoother for the flash focus sequence, and unlike the plane-wave compounding sequence, it does not exhibit the discontinuities that are characteristic of false-peak motion tracking errors~\cite{pinton2006rapid}. The frequency spectrum for the flash focus sequence (Fig. \ref{fig:Sig}{\bf b}), is consistent with theoretical predictions of odd harmonic generation, whereas the plane-wave compounding angle exhibits a second harmonic artifact which is much larger than any of the odd harmonics in its spectrum.

For a more quantitative comparison of the improvement generated by the flash focus sequence, the average spectrum was computed for two regions, first in the surrounding gelatin (Fig.~\ref{fig:fft} {\bf a}), second in the brain (Fig.~\ref{fig:fft} {\bf b}). Fig.~\ref{fig:fft} {\bf a} indicates that the two sequences generate similar results to the gelatin. The only significant improvement is a reduction of 14dB of the noise floor when the flash focus sequence was used. In contrast, in the brain region plane wave compounding underestimates the fundamental component by 7.2dB, it over estimates the second harmonic by 10.3dB and it underestimates the third harmonic by 4.4dB. Furthermore, the flash focus sequences reduces the noise floor by 14.1dB allowing visualization up to the ninth harmonic. Therefore, the flash focus sequence is a significant improvement that allows the observation and characterization of nonlinear shear waves propagated in the brain tissue.

\section*{APPENDIX C: Validation of the tracking algorithm}

The shock front of a shear wave is sharp and consequently has wide support in the spectral domain. This is challenging for tracking algorithms because the numerous high frequency harmonics are at a significantly lower amplitude than the fundamental but they define the shape and rise time of the shock front~\cite{pinton2014adaptive}. Our experimental setup does not afford opportunities for direct validation of the adaptive tracking algorithm since there is no {\it a priori} knowledge of brain motion. Therefore the adaptive tracking algorithm was validated with two simulation tools, one for nonlinear shear propagation~\cite{pinton2010nonlinear} and another for ultrasound imaging of particle displacements~\cite{Pinton2016smalldisplacement}.

\begin{figure}[!ht]
 \centering
\includegraphics[trim= 0 0 0 0, clip,width=0.48\textwidth]{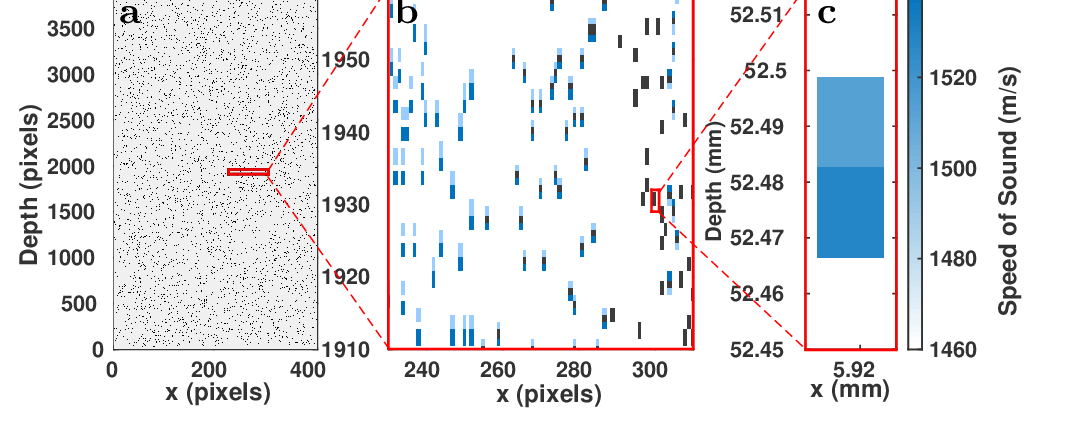}
   \caption{A realization of the acoustic map used in the finite difference time domain simulation of acoustic propagation and ultrasound imaging. The simulation domain consists of subwavelength scatterers ({\bf a}) that are shifted by a small amount determined by solution to Eq. 1 ({\bf b}) using an impedance flow method to shift the scatterers by a sub-grid displacement ({\bf c}).}
 \label{fig:stephen}
\end{figure}

First the particle displacement was determined by solving the nonlinear elastic wave equation (Eq. 1) with a Rusanov scheme \cite{rusanov1970method,pinton2010nonlinear}. This method is able to solve nonlinear hyperbolic equations with third order accuracy and low computational time. The linear and nonlinear coefficients in this equation were assigned the values that were measured in the porcine brains, $c_t=2.14$m/s, $L_A=25$mm, and $\beta=13\pm6$. The Chebyshev enveloped $75$Hz signal that was used in the experiments was also used as an initial condition to the numerical solution. An initial amplitude of $0.8$m/s was chosen since it represents the high range of what is currently achievable by our experimental setup. 

Then the output of this simulation was used to determine the position of the acoustical scatterers in a second simulation of ultrasound propagation. This second simulation tool consists of a finite difference time domain solution of the full acoustic wave equation that we have previously developed and used to model small particle displacements and to generate highly realistic ultrasound images \cite{pinton2012three,Pinton2016smalldisplacement}. Here it was used to model the field generated by the ultrasound imaging system with the same transducer geometry and transmit-receive imaging sequence. The acoustical maps generated for the simulation had the gross acoustical properties of soft tissue, i.e. an average density of $1000$kg/m$^3$, a sound speed of $1540$m/s, and an attenuation coefficient of $0.3$dB/MHz/cm. To obtain accurate speckle statistics subwavelength scatterers with a size of $39\mu m$, a concentration $>20$ scatters per resolution cell, and a random speed of sound between $1540$m/s and $1502$m/s were distributed randomly in the medium. A separate acoustical map was generated for each of the $1200$ frames and in each map the scatterers were displaced according to the time dependent displacement previously generated by the numerical solution of Eq. 1.

The macroscopic view of two scatters maps, one displaced respect to the other, is shown in Fig. \ref{fig:stephen} {\bf a}. In the mesoscopic scale ({\bf b}) the reference scatter map is shown in red and the displaced scatter map is shown in blue. Note that the displacement changes as a function of the lateral position. Subresolution displacement is represented by shifting the relative impedance of the two element that represent a single scatter (Fig. \ref{fig:stephen}{\bf c}). This is known as the impedance flow method \cite{Pinton2016smalldisplacement}.

In this scatter the displacement is less than a pixel, for that reason the displacement is imposed using an impedance flow method. A transmit-receive acoustical simulation was performed for each map to generate the raw RF data {\it in silico}. The simulated RF data was matched to experimental data by downsampling and then adding white noise to achieve the $20dB$ SNR observed in the {\it ex vivo} brain experiment. Then the RF data was beamformed using the same aperture and delay-and-sum methods that were previously used experimentally. This yielded the $1200$ frames that were finally processed with the adaptive tracking code to estimate particle motion \cite{pinton2014adaptive}.  

\begin{figure}
 \centering
\includegraphics[trim= 0 0 0 0, clip,width=0.48\textwidth]{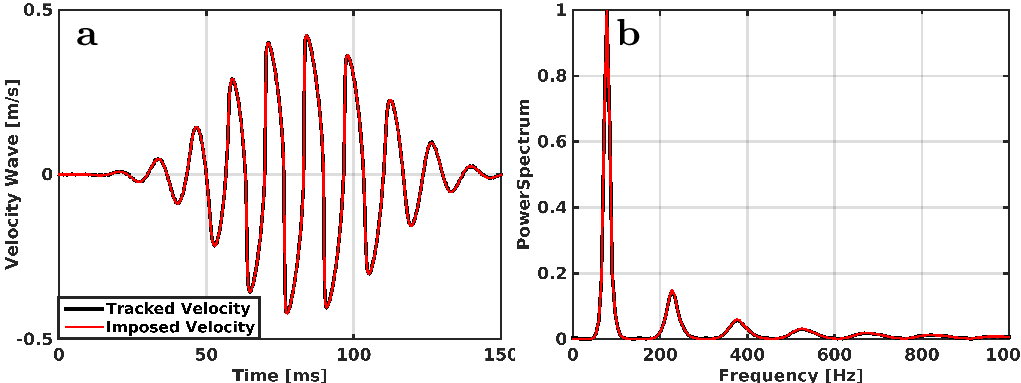}
\caption{Comparison between the imposed (red) and tracked (black) particle velocity in time domain ({\bf a}) and frequency domain ({\bf b}). The tracked velocity is obtained from the simulated and beamformed RF data, and the imposed velocity is obtained from a numerical solution of the one dimensional nonlinear elastic wave equation (Eq. 1). Note that the curves appear super-imposed.}
\label{fig:StephenSignals}
\end{figure}

The imposed shear wave calculated from Eq. 1, is shown as the red curve on the left plot in Fig. \ref{fig:StephenSignals}. It is almost indistinguishable from the ultrasound-based motion estimate shown as the black curve (the two are superimposed) and the root-mean-square difference between these two curves is $2.1$\%. The corresponding power spectrum (Fig. \ref{fig:StephenSignals}{\bf b}) is also superimposed which confirms that the ultrasound imaging sequence and adaptive tracking algorithm preserve the spectral content of the particle motion and can accurately capture the higher harmonics that are necessary for the characterization of the shock front.  

The final validation step for the tracking algorithm and its ability to represent the harmonics was performed with experiments driven at a low amplitude that was assumed to be linear ($0.03m/s$). In this experiment, shown in Fig. \ref{fig:lin}, the second and third harmonics represent less that $2.0\%$ of the amplitude of the fundamental component. Thus, the waveform does not distort nonlinearly with propagation. This not only confirms the initial assumption of linearity but also eliminates the possibility that the nonlinear behaviour observed experimentally was a tracking-algorithm based artifact.

\begin{figure}[h!]
 \centering
 \includegraphics[trim= 0 0 0 0, clip,width=0.4\textwidth]{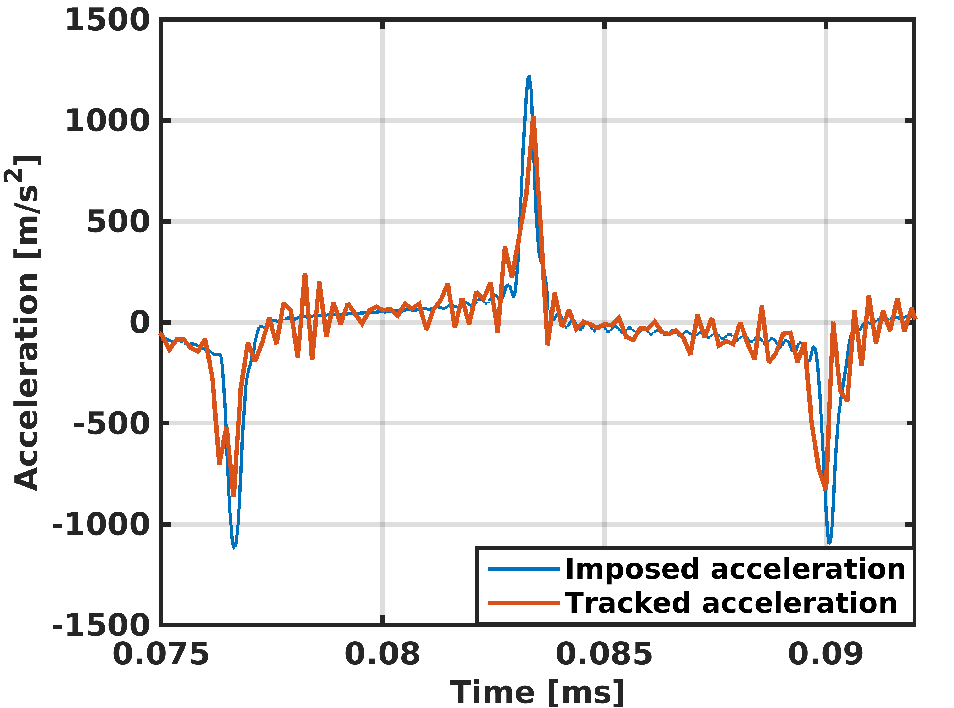}
\caption{Comparison between the imposed and tracked acceleration calculated with a Fourier-based method.}
\label{fig:acce}
\end{figure}

\section*{APPENDIX D: Differentiation Method}

Numerical differentiation, on which our estimates of acceleration from velocity rely, can be noise sensitive especially since the derivatives are required to be accurate at the steep shock front.
The acceleration, $a(t)$, was computed from the experimental data with a standard derivative in the Fourier domain \cite{oppenheim1983signals}.

To determine the expected error in acceleration the derivative of the known velocity determined by the simulation of Eq. 1 was compared to the derivative of the ultrasound-based velocity. Since the imposed velocity was calculated numerically it does not include experimental noise and it was sampled at high frequency (10 times the experimental sampling). Thus, its time derivative is robust and does not depend on the method used for differentiation. On the other hand, the experimental velocity was obtained by applying our tracking algorithm to the simulated RF data as described in the previous section. Therefore, the simulated tracked velocity has similar noise characteristics and the same sampling frequency ($6200$Hz) as the experimental data, where the actual velocity is unknown {\it a priori}. These acceleration calculations (Fig. \ref{fig:acce}) show that, even though this method increases high frequency noise proportionally to $\omega$, it preserves the acceleration peaks at the shock front and underestimates them by approximately $15\%$. This indicates that estimates of the shock front magnification factors reported in the main article are conservatively low.

\bibliographystyle{apsrev4-1}
\bibliography{pre}

\begin{thebibliography}{41}%
\makeatletter
\providecommand \@ifxundefined [1]{%
 \@ifx{#1\undefined}
}%
\providecommand \@ifnum [1]{%
 \ifnum #1\expandafter \@firstoftwo
 \else \expandafter \@secondoftwo
 \fi
}%
\providecommand \@ifx [1]{%
 \ifx #1\expandafter \@firstoftwo
 \else \expandafter \@secondoftwo
 \fi
}%
\providecommand \natexlab [1]{#1}%
\providecommand \enquote  [1]{``#1''}%
\providecommand \bibnamefont  [1]{#1}%
\providecommand \bibfnamefont [1]{#1}%
\providecommand \citenamefont [1]{#1}%
\providecommand \href@noop [0]{\@secondoftwo}%
\providecommand \href [0]{\begingroup \@sanitize@url \@href}%
\providecommand \@href[1]{\@@startlink{#1}\@@href}%
\providecommand \@@href[1]{\endgroup#1\@@endlink}%
\providecommand \@sanitize@url [0]{\catcode `\\12\catcode `\$12\catcode
  `\&12\catcode `\#12\catcode `\^12\catcode `\_12\catcode `\%12\relax}%
\providecommand \@@startlink[1]{}%
\providecommand \@@endlink[0]{}%
\providecommand \url  [0]{\begingroup\@sanitize@url \@url }%
\providecommand \@url [1]{\endgroup\@href {#1}{\urlprefix }}%
\providecommand \urlprefix  [0]{URL }%
\providecommand \Eprint [0]{\href }%
\providecommand \doibase [0]{http://dx.doi.org/}%
\providecommand \selectlanguage [0]{\@gobble}%
\providecommand \bibinfo  [0]{\@secondoftwo}%
\providecommand \bibfield  [0]{\@secondoftwo}%
\providecommand \translation [1]{[#1]}%
\providecommand \BibitemOpen [0]{}%
\providecommand \bibitemStop [0]{}%
\providecommand \bibitemNoStop [0]{.\EOS\space}%
\providecommand \EOS [0]{\spacefactor3000\relax}%
\providecommand \BibitemShut  [1]{\csname bibitem#1\endcsname}%
\let\auto@bib@innerbib\@empty
%</preamble>
\bibitem [{\citenamefont {Rimel}\ \emph {et~al.}(1981)\citenamefont {Rimel},
  \citenamefont {Giordani}, \citenamefont {Barth}, \citenamefont {Boll},\ and\
  \citenamefont {Jane}}]{rimel1981disability}%
  \BibitemOpen
  \bibfield  {author} {\bibinfo {author} {\bibfnamefont {R.~W.}\ \bibnamefont
  {Rimel}}, \bibinfo {author} {\bibfnamefont {B.}~\bibnamefont {Giordani}},
  \bibinfo {author} {\bibfnamefont {J.~T.}\ \bibnamefont {Barth}}, \bibinfo
  {author} {\bibfnamefont {T.~J.}\ \bibnamefont {Boll}}, \ and\ \bibinfo
  {author} {\bibfnamefont {J.~A.}\ \bibnamefont {Jane}},\ }\href@noop {}
  {\bibfield  {journal} {\bibinfo  {journal} {Neurosurgery}\ }\textbf {\bibinfo
  {volume} {9}},\ \bibinfo {pages} {221} (\bibinfo {year} {1981})}\BibitemShut
  {NoStop}%
\bibitem [{\citenamefont {Rowson}\ and\ \citenamefont
  {Duma}(2013)}]{rowson2013brain}%
  \BibitemOpen
  \bibfield  {author} {\bibinfo {author} {\bibfnamefont {S.}~\bibnamefont
  {Rowson}}\ and\ \bibinfo {author} {\bibfnamefont {S.~M.}\ \bibnamefont
  {Duma}},\ }\href@noop {} {\bibfield  {journal} {\bibinfo  {journal} {Ann.
  Biomed. Eng.}\ }\textbf {\bibinfo {volume} {41}},\ \bibinfo {pages} {873}
  (\bibinfo {year} {2013})}\BibitemShut {NoStop}%
\bibitem [{\citenamefont {Broglio}\ \emph {et~al.}(2009)\citenamefont
  {Broglio}, \citenamefont {Sosnoff}, \citenamefont {Shin}, \citenamefont {He},
  \citenamefont {Alcaraz},\ and\ \citenamefont {Zimmerman}}]{Broglio2009}%
  \BibitemOpen
  \bibfield  {author} {\bibinfo {author} {\bibfnamefont {S.~P.}\ \bibnamefont
  {Broglio}}, \bibinfo {author} {\bibfnamefont {J.~J.}\ \bibnamefont
  {Sosnoff}}, \bibinfo {author} {\bibfnamefont {S.}~\bibnamefont {Shin}},
  \bibinfo {author} {\bibfnamefont {X.}~\bibnamefont {He}}, \bibinfo {author}
  {\bibfnamefont {C.}~\bibnamefont {Alcaraz}}, \ and\ \bibinfo {author}
  {\bibfnamefont {J.}~\bibnamefont {Zimmerman}},\ }\href@noop {} {\bibfield
  {journal} {\bibinfo  {journal} {J. Athl. Training}\ }\textbf {\bibinfo
  {volume} {44}},\ \bibinfo {pages} {342} (\bibinfo {year} {2009})}\BibitemShut
  {NoStop}%
\bibitem [{\citenamefont {Crisco}\ \emph {et~al.}(2011)\citenamefont {Crisco},
  \citenamefont {Wilcox}, \citenamefont {Beckwith}, \citenamefont {Chu},
  \citenamefont {Duhaime}, \citenamefont {Rowson}, \citenamefont {Duma},
  \citenamefont {Maerlender}, \citenamefont {McAllister},\ and\ \citenamefont
  {Greenwald}}]{Crisco20112673}%
  \BibitemOpen
  \bibfield  {author} {\bibinfo {author} {\bibfnamefont {J.~J.}\ \bibnamefont
  {Crisco}}, \bibinfo {author} {\bibfnamefont {B.~J.}\ \bibnamefont {Wilcox}},
  \bibinfo {author} {\bibfnamefont {J.~G.}\ \bibnamefont {Beckwith}}, \bibinfo
  {author} {\bibfnamefont {J.~J.}\ \bibnamefont {Chu}}, \bibinfo {author}
  {\bibfnamefont {A.-C.}\ \bibnamefont {Duhaime}}, \bibinfo {author}
  {\bibfnamefont {S.}~\bibnamefont {Rowson}}, \bibinfo {author} {\bibfnamefont
  {S.~M.}\ \bibnamefont {Duma}}, \bibinfo {author} {\bibfnamefont {A.~C.}\
  \bibnamefont {Maerlender}}, \bibinfo {author} {\bibfnamefont {T.~W.}\
  \bibnamefont {McAllister}}, \ and\ \bibinfo {author} {\bibfnamefont {R.~M.}\
  \bibnamefont {Greenwald}},\ }\href {\doibase
  http://dx.doi.org/10.1016/j.jbiomech.2011.08.003} {\bibfield  {journal}
  {\bibinfo  {journal} {J. Biomech.}\ }\textbf {\bibinfo {volume} {44}},\
  \bibinfo {pages} {2673 } (\bibinfo {year} {2011})}\BibitemShut {NoStop}%
\bibitem [{\citenamefont {Daniel}\ \emph {et~al.}(2012)\citenamefont {Daniel},
  \citenamefont {Rowson},\ and\ \citenamefont {Duma}}]{Daniel2012}%
  \BibitemOpen
  \bibfield  {author} {\bibinfo {author} {\bibfnamefont {R.~W.}\ \bibnamefont
  {Daniel}}, \bibinfo {author} {\bibfnamefont {S.}~\bibnamefont {Rowson}}, \
  and\ \bibinfo {author} {\bibfnamefont {S.~M.}\ \bibnamefont {Duma}},\ }\href
  {\doibase 10.1007/s10439-012-0530-7} {\bibfield  {journal} {\bibinfo
  {journal} {Ann. Biomed. Eng.}\ }\textbf {\bibinfo {volume} {40}},\ \bibinfo
  {pages} {976} (\bibinfo {year} {2012})}\BibitemShut {NoStop}%
\bibitem [{\citenamefont {Laksari}\ \emph {et~al.}(2015)\citenamefont
  {Laksari}, \citenamefont {Wu}, \citenamefont {Kurt}, \citenamefont {Kuo},\
  and\ \citenamefont {Camarillo}}]{Laksari20150331}%
  \BibitemOpen
  \bibfield  {author} {\bibinfo {author} {\bibfnamefont {K.}~\bibnamefont
  {Laksari}}, \bibinfo {author} {\bibfnamefont {L.~C.}\ \bibnamefont {Wu}},
  \bibinfo {author} {\bibfnamefont {M.}~\bibnamefont {Kurt}}, \bibinfo {author}
  {\bibfnamefont {C.}~\bibnamefont {Kuo}}, \ and\ \bibinfo {author}
  {\bibfnamefont {D.~C.}\ \bibnamefont {Camarillo}},\ }\href@noop {} {\bibfield
   {journal} {\bibinfo  {journal} {J. R. Soc. Interface}\ }\textbf {\bibinfo
  {volume} {12}} (\bibinfo {year} {2015})}\BibitemShut {NoStop}%
\bibitem [{\citenamefont {Wright}\ and\ \citenamefont
  {Ramesh}(2012)}]{Wright2012}%
  \BibitemOpen
  \bibfield  {author} {\bibinfo {author} {\bibfnamefont {R.~M.}\ \bibnamefont
  {Wright}}\ and\ \bibinfo {author} {\bibfnamefont {K.~T.}\ \bibnamefont
  {Ramesh}},\ }\href {\doibase 10.1007/s10237-011-0307-1} {\bibfield  {journal}
  {\bibinfo  {journal} {Biomech. Model. Mechan.}\ }\textbf {\bibinfo {volume}
  {11}},\ \bibinfo {pages} {245} (\bibinfo {year} {2012})}\BibitemShut
  {NoStop}%
\bibitem [{\citenamefont {Guskiewicz}\ \emph {et~al.}(2007)\citenamefont
  {Guskiewicz}, \citenamefont {Mihalik}, \citenamefont {Shankar}, \citenamefont
  {Marshall}, \citenamefont {Crowell}, \citenamefont {Oliaro}, \citenamefont
  {Ciocca},\ and\ \citenamefont {Hooker}}]{guskiewicz2007measurement}%
  \BibitemOpen
  \bibfield  {author} {\bibinfo {author} {\bibfnamefont {K.~M.}\ \bibnamefont
  {Guskiewicz}}, \bibinfo {author} {\bibfnamefont {J.~P.}\ \bibnamefont
  {Mihalik}}, \bibinfo {author} {\bibfnamefont {V.}~\bibnamefont {Shankar}},
  \bibinfo {author} {\bibfnamefont {S.~W.}\ \bibnamefont {Marshall}}, \bibinfo
  {author} {\bibfnamefont {D.~H.}\ \bibnamefont {Crowell}}, \bibinfo {author}
  {\bibfnamefont {S.~M.}\ \bibnamefont {Oliaro}}, \bibinfo {author}
  {\bibfnamefont {M.~F.}\ \bibnamefont {Ciocca}}, \ and\ \bibinfo {author}
  {\bibfnamefont {D.~N.}\ \bibnamefont {Hooker}},\ }\href@noop {} {\bibfield
  {journal} {\bibinfo  {journal} {Neurosurgery}\ }\textbf {\bibinfo {volume}
  {61}},\ \bibinfo {pages} {1244} (\bibinfo {year} {2007})}\BibitemShut
  {NoStop}%
\bibitem [{\citenamefont {Bayly}\ \emph {et~al.}(2005)\citenamefont {Bayly},
  \citenamefont {Cohen}, \citenamefont {Leister}, \citenamefont {Ajo},
  \citenamefont {Leuthardt},\ and\ \citenamefont
  {Genin}}]{bayly2005deformation}%
  \BibitemOpen
  \bibfield  {author} {\bibinfo {author} {\bibfnamefont {P.}~\bibnamefont
  {Bayly}}, \bibinfo {author} {\bibfnamefont {T.}~\bibnamefont {Cohen}},
  \bibinfo {author} {\bibfnamefont {E.}~\bibnamefont {Leister}}, \bibinfo
  {author} {\bibfnamefont {D.}~\bibnamefont {Ajo}}, \bibinfo {author}
  {\bibfnamefont {E.}~\bibnamefont {Leuthardt}}, \ and\ \bibinfo {author}
  {\bibfnamefont {G.}~\bibnamefont {Genin}},\ }\href@noop {} {\bibfield
  {journal} {\bibinfo  {journal} {J. Neurotraum.}\ }\textbf {\bibinfo {volume}
  {22}},\ \bibinfo {pages} {845} (\bibinfo {year} {2005})}\BibitemShut
  {NoStop}%
\bibitem [{\citenamefont {Moss}\ \emph {et~al.}(2009)\citenamefont {Moss},
  \citenamefont {King},\ and\ \citenamefont {Blackman}}]{MossBrain}%
  \BibitemOpen
  \bibfield  {author} {\bibinfo {author} {\bibfnamefont {W.~C.}\ \bibnamefont
  {Moss}}, \bibinfo {author} {\bibfnamefont {M.~J.}\ \bibnamefont {King}}, \
  and\ \bibinfo {author} {\bibfnamefont {E.~G.}\ \bibnamefont {Blackman}},\
  }\href {\doibase 10.1103/PhysRevLett.103.108702} {\bibfield  {journal}
  {\bibinfo  {journal} {Phys. Rev. Lett.}\ }\textbf {\bibinfo {volume} {103}},\
  \bibinfo {pages} {108702} (\bibinfo {year} {2009})}\BibitemShut {NoStop}%
\bibitem [{\citenamefont {Goriely}\ \emph {et~al.}(2016)\citenamefont
  {Goriely}, \citenamefont {Weickenmeier},\ and\ \citenamefont
  {Kuhl}}]{GorielyBrain}%
  \BibitemOpen
  \bibfield  {author} {\bibinfo {author} {\bibfnamefont {A.}~\bibnamefont
  {Goriely}}, \bibinfo {author} {\bibfnamefont {J.}~\bibnamefont
  {Weickenmeier}}, \ and\ \bibinfo {author} {\bibfnamefont {E.}~\bibnamefont
  {Kuhl}},\ }\href {\doibase 10.1103/PhysRevLett.117.138001} {\bibfield
  {journal} {\bibinfo  {journal} {Phys. Rev. Lett.}\ }\textbf {\bibinfo
  {volume} {117}},\ \bibinfo {pages} {138001} (\bibinfo {year}
  {2016})}\BibitemShut {NoStop}%
\bibitem [{\citenamefont {Darvish}\ and\ \citenamefont
  {Crandall}(2001)}]{darvish2001nonlinear}%
  \BibitemOpen
  \bibfield  {author} {\bibinfo {author} {\bibfnamefont {K.}~\bibnamefont
  {Darvish}}\ and\ \bibinfo {author} {\bibfnamefont {J.}~\bibnamefont
  {Crandall}},\ }\href@noop {} {\bibfield  {journal} {\bibinfo  {journal} {Med.
  Eng. Phys.}\ }\textbf {\bibinfo {volume} {23}},\ \bibinfo {pages} {633}
  (\bibinfo {year} {2001})}\BibitemShut {NoStop}%
\bibitem [{\citenamefont {Donnelly}(1997)}]{Donnelly1997}%
  \BibitemOpen
  \bibfield  {author} {\bibinfo {author} {\bibfnamefont {B.~R.}\ \bibnamefont
  {Donnelly}},\ }\href {\doibase http://dx.doi.org/10.1115/1.2798289}
  {\bibfield  {journal} {\bibinfo  {journal} {J. Biomech. Eng.}\ }\textbf
  {\bibinfo {volume} {119}},\ \bibinfo {pages} {423 } (\bibinfo {year}
  {1997})}\BibitemShut {NoStop}%
\bibitem [{\citenamefont {Margulies}\ and\ \citenamefont
  {Thibault}(1992)}]{margulies1992}%
  \BibitemOpen
  \bibfield  {author} {\bibinfo {author} {\bibfnamefont {S.~S.}\ \bibnamefont
  {Margulies}}\ and\ \bibinfo {author} {\bibfnamefont {L.~E.}\ \bibnamefont
  {Thibault}},\ }\href {\doibase
  http://dx.doi.org/10.1016/0021-9290(92)90231-O} {\bibfield  {journal}
  {\bibinfo  {journal} {J. Biomech.}\ }\textbf {\bibinfo {volume} {25}},\
  \bibinfo {pages} {917 } (\bibinfo {year} {1992})}\BibitemShut {NoStop}%
\bibitem [{\citenamefont {Sullivan}\ \emph {et~al.}(2015)\citenamefont
  {Sullivan}, \citenamefont {Eucker}, \citenamefont {Gabrieli}, \citenamefont
  {Bradfield}, \citenamefont {Coats}, \citenamefont {Maltese}, \citenamefont
  {Lee}, \citenamefont {Smith},\ and\ \citenamefont
  {Margulies}}]{Sullivan2015}%
  \BibitemOpen
  \bibfield  {author} {\bibinfo {author} {\bibfnamefont {S.}~\bibnamefont
  {Sullivan}}, \bibinfo {author} {\bibfnamefont {S.~A.}\ \bibnamefont
  {Eucker}}, \bibinfo {author} {\bibfnamefont {D.}~\bibnamefont {Gabrieli}},
  \bibinfo {author} {\bibfnamefont {C.}~\bibnamefont {Bradfield}}, \bibinfo
  {author} {\bibfnamefont {B.}~\bibnamefont {Coats}}, \bibinfo {author}
  {\bibfnamefont {M.~R.}\ \bibnamefont {Maltese}}, \bibinfo {author}
  {\bibfnamefont {J.}~\bibnamefont {Lee}}, \bibinfo {author} {\bibfnamefont
  {C.}~\bibnamefont {Smith}}, \ and\ \bibinfo {author} {\bibfnamefont {S.~S.}\
  \bibnamefont {Margulies}},\ }\href@noop {} {\bibfield  {journal} {\bibinfo
  {journal} {Biomech. Model. Mechan.}\ }\textbf {\bibinfo {volume} {14}},\
  \bibinfo {pages} {877} (\bibinfo {year} {2015})}\BibitemShut {NoStop}%
\bibitem [{\citenamefont {Hamhaber}\ \emph {et~al.}(2010)\citenamefont
  {Hamhaber}, \citenamefont {Klatt}, \citenamefont {Papazoglou}, \citenamefont
  {Hollmann}, \citenamefont {Stadler}, \citenamefont {Sack}, \citenamefont
  {Bernarding},\ and\ \citenamefont {Braun}}]{hamhaber2010vivo}%
  \BibitemOpen
  \bibfield  {author} {\bibinfo {author} {\bibfnamefont {U.}~\bibnamefont
  {Hamhaber}}, \bibinfo {author} {\bibfnamefont {D.}~\bibnamefont {Klatt}},
  \bibinfo {author} {\bibfnamefont {S.}~\bibnamefont {Papazoglou}}, \bibinfo
  {author} {\bibfnamefont {M.}~\bibnamefont {Hollmann}}, \bibinfo {author}
  {\bibfnamefont {J.}~\bibnamefont {Stadler}}, \bibinfo {author} {\bibfnamefont
  {I.}~\bibnamefont {Sack}}, \bibinfo {author} {\bibfnamefont {J.}~\bibnamefont
  {Bernarding}}, \ and\ \bibinfo {author} {\bibfnamefont {J.}~\bibnamefont
  {Braun}},\ }\href@noop {} {\bibfield  {journal} {\bibinfo  {journal} {J.
  Magn. Reson. Imaging}\ }\textbf {\bibinfo {volume} {32}},\ \bibinfo {pages}
  {577} (\bibinfo {year} {2010})}\BibitemShut {NoStop}%
\bibitem [{\citenamefont {Johnson}\ \emph {et~al.}(2013)\citenamefont
  {Johnson}, \citenamefont {McGarry}, \citenamefont {Houten}, \citenamefont
  {Weaver}, \citenamefont {Paulsen}, \citenamefont {Sutton},\ and\
  \citenamefont {Georgiadis}}]{johnson2013magnetic}%
  \BibitemOpen
  \bibfield  {author} {\bibinfo {author} {\bibfnamefont {C.~L.}\ \bibnamefont
  {Johnson}}, \bibinfo {author} {\bibfnamefont {M.~D.}\ \bibnamefont
  {McGarry}}, \bibinfo {author} {\bibfnamefont {E.~E.}\ \bibnamefont {Houten}},
  \bibinfo {author} {\bibfnamefont {J.~B.}\ \bibnamefont {Weaver}}, \bibinfo
  {author} {\bibfnamefont {K.~D.}\ \bibnamefont {Paulsen}}, \bibinfo {author}
  {\bibfnamefont {B.~P.}\ \bibnamefont {Sutton}}, \ and\ \bibinfo {author}
  {\bibfnamefont {J.~G.}\ \bibnamefont {Georgiadis}},\ }\href@noop {}
  {\bibfield  {journal} {\bibinfo  {journal} {Magn. Reson. Med.}\ }\textbf
  {\bibinfo {volume} {70}},\ \bibinfo {pages} {404} (\bibinfo {year}
  {2013})}\BibitemShut {NoStop}%
\bibitem [{\citenamefont {Razani}\ \emph {et~al.}(2012)\citenamefont {Razani},
  \citenamefont {Mariampillai}, \citenamefont {Sun}, \citenamefont {Luk},
  \citenamefont {Yang},\ and\ \citenamefont {Kolios}}]{Razani2012}%
  \BibitemOpen
  \bibfield  {author} {\bibinfo {author} {\bibfnamefont {M.}~\bibnamefont
  {Razani}}, \bibinfo {author} {\bibfnamefont {A.}~\bibnamefont
  {Mariampillai}}, \bibinfo {author} {\bibfnamefont {C.}~\bibnamefont {Sun}},
  \bibinfo {author} {\bibfnamefont {T.~W.}\ \bibnamefont {Luk}}, \bibinfo
  {author} {\bibfnamefont {V.~X.}\ \bibnamefont {Yang}}, \ and\ \bibinfo
  {author} {\bibfnamefont {M.~C.}\ \bibnamefont {Kolios}},\ }\href@noop {}
  {\bibfield  {journal} {\bibinfo  {journal} {Biomedical optics express}\
  }\textbf {\bibinfo {volume} {3}},\ \bibinfo {pages} {972} (\bibinfo {year}
  {2012})}\BibitemShut {NoStop}%
\bibitem [{\citenamefont {Song}\ \emph {et~al.}(2013)\citenamefont {Song},
  \citenamefont {Huang},\ and\ \citenamefont {Wang}}]{Song2013}%
  \BibitemOpen
  \bibfield  {author} {\bibinfo {author} {\bibfnamefont {S.}~\bibnamefont
  {Song}}, \bibinfo {author} {\bibfnamefont {Z.}~\bibnamefont {Huang}}, \ and\
  \bibinfo {author} {\bibfnamefont {R.~K.}\ \bibnamefont {Wang}},\ }\href@noop
  {} {\bibfield  {journal} {\bibinfo  {journal} {J. Biomed. Opt.}\ }\textbf
  {\bibinfo {volume} {18}},\ \bibinfo {pages} {121505} (\bibinfo {year}
  {2013})}\BibitemShut {NoStop}%
\bibitem [{\citenamefont {Razani}\ \emph {et~al.}(2014)\citenamefont {Razani},
  \citenamefont {Luk}, \citenamefont {Mariampillai}, \citenamefont {Siegler},
  \citenamefont {Kiehl}, \citenamefont {Kolios},\ and\ \citenamefont
  {Yang}}]{Razani2014}%
  \BibitemOpen
  \bibfield  {author} {\bibinfo {author} {\bibfnamefont {M.}~\bibnamefont
  {Razani}}, \bibinfo {author} {\bibfnamefont {T.~W.}\ \bibnamefont {Luk}},
  \bibinfo {author} {\bibfnamefont {A.}~\bibnamefont {Mariampillai}}, \bibinfo
  {author} {\bibfnamefont {P.}~\bibnamefont {Siegler}}, \bibinfo {author}
  {\bibfnamefont {T.-R.}\ \bibnamefont {Kiehl}}, \bibinfo {author}
  {\bibfnamefont {M.~C.}\ \bibnamefont {Kolios}}, \ and\ \bibinfo {author}
  {\bibfnamefont {V.~X.}\ \bibnamefont {Yang}},\ }\href@noop {} {\bibfield
  {journal} {\bibinfo  {journal} {Biomedical optics express}\ }\textbf
  {\bibinfo {volume} {5}},\ \bibinfo {pages} {895} (\bibinfo {year}
  {2014})}\BibitemShut {NoStop}%
\bibitem [{\citenamefont {Park}\ \emph {et~al.}(2015)\citenamefont {Park},
  \citenamefont {Sun},\ and\ \citenamefont {Cui}}]{Park28072015}%
  \BibitemOpen
  \bibfield  {author} {\bibinfo {author} {\bibfnamefont {J.-H.}\ \bibnamefont
  {Park}}, \bibinfo {author} {\bibfnamefont {W.}~\bibnamefont {Sun}}, \ and\
  \bibinfo {author} {\bibfnamefont {M.}~\bibnamefont {Cui}},\ }\href@noop {}
  {\bibfield  {journal} {\bibinfo  {journal} {P. Natl. Acad. Sci. USA}\
  }\textbf {\bibinfo {volume} {112}},\ \bibinfo {pages} {9236} (\bibinfo {year}
  {2015})}\BibitemShut {NoStop}%
\bibitem [{\citenamefont {Catheline}\ \emph {et~al.}(2003)\citenamefont
  {Catheline}, \citenamefont {Gennisson}, \citenamefont {Tanter},\ and\
  \citenamefont {Fink}}]{catheline2003observation}%
  \BibitemOpen
  \bibfield  {author} {\bibinfo {author} {\bibfnamefont {S.}~\bibnamefont
  {Catheline}}, \bibinfo {author} {\bibfnamefont {J.-L.}\ \bibnamefont
  {Gennisson}}, \bibinfo {author} {\bibfnamefont {M.}~\bibnamefont {Tanter}}, \
  and\ \bibinfo {author} {\bibfnamefont {M.}~\bibnamefont {Fink}},\ }\href@noop
  {} {\bibfield  {journal} {\bibinfo  {journal} {Phys. Rev. Lett.}\ }\textbf
  {\bibinfo {volume} {91}},\ \bibinfo {pages} {164301} (\bibinfo {year}
  {2003})}\BibitemShut {NoStop}%
\bibitem [{\citenamefont {Tanter}\ \emph {et~al.}(2002)\citenamefont {Tanter},
  \citenamefont {Bercoff}, \citenamefont {Sandrin},\ and\ \citenamefont
  {Fink}}]{tanter2002ultrafast}%
  \BibitemOpen
  \bibfield  {author} {\bibinfo {author} {\bibfnamefont {M.}~\bibnamefont
  {Tanter}}, \bibinfo {author} {\bibfnamefont {J.}~\bibnamefont {Bercoff}},
  \bibinfo {author} {\bibfnamefont {L.}~\bibnamefont {Sandrin}}, \ and\
  \bibinfo {author} {\bibfnamefont {M.}~\bibnamefont {Fink}},\ }\href@noop {}
  {\bibfield  {journal} {\bibinfo  {journal} {IEEE T. Ultrason. Ferr.}\
  }\textbf {\bibinfo {volume} {49}},\ \bibinfo {pages} {1363} (\bibinfo {year}
  {2002})}\BibitemShut {NoStop}%
\bibitem [{\citenamefont {Pinton}\ \emph {et~al.}(2014)\citenamefont {Pinton},
  \citenamefont {Gennisson}, \citenamefont {Tanter},\ and\ \citenamefont
  {Coulouvrat}}]{pinton2014adaptive}%
  \BibitemOpen
  \bibfield  {author} {\bibinfo {author} {\bibfnamefont {G.}~\bibnamefont
  {Pinton}}, \bibinfo {author} {\bibfnamefont {J.-L.}\ \bibnamefont
  {Gennisson}}, \bibinfo {author} {\bibfnamefont {M.}~\bibnamefont {Tanter}}, \
  and\ \bibinfo {author} {\bibfnamefont {F.}~\bibnamefont {Coulouvrat}},\
  }\href@noop {} {\bibfield  {journal} {\bibinfo  {journal} {IEEE T. Ultrason.
  Ferr.}\ }\textbf {\bibinfo {volume} {61}},\ \bibinfo {pages} {1489} (\bibinfo
  {year} {2014})}\BibitemShut {NoStop}%
\bibitem [{sup()}]{sup}%
  \BibitemOpen
  \href@noop {} {\emph {\bibinfo {title} {See Supplemental Material at [URL
  will be inserted by publisher] for: a Movie of the ultrasound Bmode images. b
  Movie of the particule velocity and movie. c Movie of the particle
  acceleration of the brain.}}}\BibitemShut {Stop}%
\bibitem [{\citenamefont {Duck}(1990)}]{duck1990physical}%
  \BibitemOpen
  \bibfield  {author} {\bibinfo {author} {\bibfnamefont {F.}~\bibnamefont
  {Duck}},\ }\href@noop {} {\emph {\bibinfo {title} {{Physical properties of
  tissue: a comprehensive reference book}}}}\ (\bibinfo  {publisher} {Academic
  Press, London, UK},\ \bibinfo {year} {1990})\BibitemShut {NoStop}%
\bibitem [{\citenamefont {Giammarinaro}\ \emph {et~al.}(2014)\citenamefont
  {Giammarinaro}, \citenamefont {Coulouvrat},\ and\ \citenamefont
  {Pinton}}]{giammarinaro2014numerical}%
  \BibitemOpen
  \bibfield  {author} {\bibinfo {author} {\bibfnamefont {B.}~\bibnamefont
  {Giammarinaro}}, \bibinfo {author} {\bibfnamefont {F.}~\bibnamefont
  {Coulouvrat}}, \ and\ \bibinfo {author} {\bibfnamefont {G.}~\bibnamefont
  {Pinton}},\ }\href@noop {} {\bibfield  {journal} {\bibinfo  {journal}
  {Physics in Medecine and Biology (Submitted)}\ } (\bibinfo {year}
  {2014})}\BibitemShut {NoStop}%
\bibitem [{\citenamefont {Greenwald}\ \emph {et~al.}(2008)\citenamefont
  {Greenwald}, \citenamefont {Gwin}, \citenamefont {Chu},\ and\ \citenamefont
  {Crisco}}]{greenwald2008head}%
  \BibitemOpen
  \bibfield  {author} {\bibinfo {author} {\bibfnamefont {R.~M.}\ \bibnamefont
  {Greenwald}}, \bibinfo {author} {\bibfnamefont {J.~T.}\ \bibnamefont {Gwin}},
  \bibinfo {author} {\bibfnamefont {J.~J.}\ \bibnamefont {Chu}}, \ and\
  \bibinfo {author} {\bibfnamefont {J.~J.}\ \bibnamefont {Crisco}},\
  }\href@noop {} {\bibfield  {journal} {\bibinfo  {journal} {Neurosurgery}\
  }\textbf {\bibinfo {volume} {62}},\ \bibinfo {pages} {789} (\bibinfo {year}
  {2008})}\BibitemShut {NoStop}%
\bibitem [{\citenamefont {Montaldo}\ \emph {et~al.}(2009)\citenamefont
  {Montaldo}, \citenamefont {Tanter}, \citenamefont {Bercoff}, \citenamefont
  {Benech},\ and\ \citenamefont {Fink}}]{montaldo2009flashangles}%
  \BibitemOpen
  \bibfield  {author} {\bibinfo {author} {\bibfnamefont {G.}~\bibnamefont
  {Montaldo}}, \bibinfo {author} {\bibfnamefont {M.}~\bibnamefont {Tanter}},
  \bibinfo {author} {\bibfnamefont {J.}~\bibnamefont {Bercoff}}, \bibinfo
  {author} {\bibfnamefont {N.}~\bibnamefont {Benech}}, \ and\ \bibinfo {author}
  {\bibfnamefont {M.}~\bibnamefont {Fink}},\ }\href {\doibase
  10.1109/TUFFC.2009.1067} {\bibfield  {journal} {\bibinfo  {journal} {IEEE T.
  Ultrason. Ferr.}\ }\textbf {\bibinfo {volume} {56}},\ \bibinfo {pages} {489}
  (\bibinfo {year} {2009})}\BibitemShut {NoStop}%
\bibitem [{\citenamefont {Pinton}\ \emph {et~al.}(2011)\citenamefont {Pinton},
  \citenamefont {Trahey},\ and\ \citenamefont {Dahl}}]{pinton2011sources}%
  \BibitemOpen
  \bibfield  {author} {\bibinfo {author} {\bibfnamefont {G.~F.}\ \bibnamefont
  {Pinton}}, \bibinfo {author} {\bibfnamefont {G.~E.}\ \bibnamefont {Trahey}},
  \ and\ \bibinfo {author} {\bibfnamefont {J.~J.}\ \bibnamefont {Dahl}},\
  }\href@noop {} {\bibfield  {journal} {\bibinfo  {journal} {IEEE T. Ultrason.
  Ferr.}\ }\textbf {\bibinfo {volume} {58}},\ \bibinfo {pages} {754} (\bibinfo
  {year} {2011})}\BibitemShut {NoStop}%
\bibitem [{\citenamefont {Pinton}(2012)}]{pinton2012three}%
  \BibitemOpen
  \bibfield  {author} {\bibinfo {author} {\bibfnamefont {G.}~\bibnamefont
  {Pinton}},\ }in\ \href@noop {} {\emph {\bibinfo {booktitle} {IEEE
  International Ultrasonics Symposium Proceeding, Dresden, Germany}}}\
  (\bibinfo {year} {2012})\BibitemShut {NoStop}%
\bibitem [{\citenamefont {Pinton}(2017)}]{Pinton2016smalldisplacement}%
  \BibitemOpen
  \bibfield  {author} {\bibinfo {author} {\bibfnamefont {G.}~\bibnamefont
  {Pinton}},\ }\href@noop {} {\bibfield  {journal} {\bibinfo  {journal} {IEEE
  T. Ultrason. Ferr. (Accepted)}\ }\textbf {\bibinfo {volume} {64}},\ \bibinfo
  {pages} {537} (\bibinfo {year} {2017})}\BibitemShut {NoStop}%
\bibitem [{\citenamefont {Landau}\ and\ \citenamefont
  {Lifshitz}(1960)}]{landau1960theory}%
  \BibitemOpen
  \bibfield  {author} {\bibinfo {author} {\bibfnamefont {L.~D.}\ \bibnamefont
  {Landau}}\ and\ \bibinfo {author} {\bibfnamefont {E.~M.}\ \bibnamefont
  {Lifshitz}},\ }\href@noop {} {\emph {\bibinfo {title} {{Theory of Elasticity:
  Vol. 7 of Course of Theoretical Physics}}}}\ (\bibinfo {year}
  {1960})\BibitemShut {NoStop}%
\bibitem [{\citenamefont {Wochner}\ \emph {et~al.}(2008)\citenamefont
  {Wochner}, \citenamefont {Hamilton}, \citenamefont {Ilinksii},\ and\
  \citenamefont {Zabolotskaya}}]{Wochner2008}%
  \BibitemOpen
  \bibfield  {author} {\bibinfo {author} {\bibfnamefont {M.}~\bibnamefont
  {Wochner}}, \bibinfo {author} {\bibfnamefont {M.}~\bibnamefont {Hamilton}},
  \bibinfo {author} {\bibfnamefont {Y.}~\bibnamefont {Ilinksii}}, \ and\
  \bibinfo {author} {\bibfnamefont {E.}~\bibnamefont {Zabolotskaya}},\
  }\href@noop {} {\bibfield  {journal} {\bibinfo  {journal} {J. Acoust. Soc.
  Am.}\ }\textbf {\bibinfo {volume} {123}},\ \bibinfo {pages} {2488} (\bibinfo
  {year} {2008})}\BibitemShut {NoStop}%
\bibitem [{\citenamefont {Zabolotskaya}\ \emph {et~al.}(2004)\citenamefont
  {Zabolotskaya}, \citenamefont {Hamilton}, \citenamefont {Ilinskii},\ and\
  \citenamefont {Meegan}}]{zabolotskaya2004modeling}%
  \BibitemOpen
  \bibfield  {author} {\bibinfo {author} {\bibfnamefont {E.}~\bibnamefont
  {Zabolotskaya}}, \bibinfo {author} {\bibfnamefont {M.}~\bibnamefont
  {Hamilton}}, \bibinfo {author} {\bibfnamefont {Y.}~\bibnamefont {Ilinskii}},
  \ and\ \bibinfo {author} {\bibfnamefont {G.}~\bibnamefont {Meegan}},\
  }\href@noop {} {\bibfield  {journal} {\bibinfo  {journal} {J. Acoust. Soc.
  Am.}\ }\textbf {\bibinfo {volume} {116}},\ \bibinfo {pages} {2807} (\bibinfo
  {year} {2004})}\BibitemShut {NoStop}%
\bibitem [{\citenamefont {Pinton}\ \emph {et~al.}(2010)\citenamefont {Pinton},
  \citenamefont {Coulouvrat}, \citenamefont {Gennisson},\ and\ \citenamefont
  {Tanter}}]{pinton2010nonlinear}%
  \BibitemOpen
  \bibfield  {author} {\bibinfo {author} {\bibfnamefont {G.}~\bibnamefont
  {Pinton}}, \bibinfo {author} {\bibfnamefont {F.}~\bibnamefont {Coulouvrat}},
  \bibinfo {author} {\bibfnamefont {J.}~\bibnamefont {Gennisson}}, \ and\
  \bibinfo {author} {\bibfnamefont {M.}~\bibnamefont {Tanter}},\ }\href@noop {}
  {\bibfield  {journal} {\bibinfo  {journal} {J. Acoust. Soc. Am.}\ }\textbf
  {\bibinfo {volume} {127}},\ \bibinfo {pages} {683} (\bibinfo {year}
  {2010})}\BibitemShut {NoStop}%
\bibitem [{\citenamefont {Whitham}(1974)}]{whitham1974}%
  \BibitemOpen
  \bibfield  {author} {\bibinfo {author} {\bibfnamefont {G.~B.}\ \bibnamefont
  {Whitham}},\ }\href@noop {} {\emph {\bibinfo {title} {Linear and nonlinear
  waves}}}\ (\bibinfo  {publisher} {Wiley-Interscience Series of Texts,
  Monographs and Tracts},\ \bibinfo {year} {1974})\BibitemShut {NoStop}%
\bibitem [{\citenamefont {Burgers}(1948)}]{burgers1948mathematical}%
  \BibitemOpen
  \bibfield  {author} {\bibinfo {author} {\bibfnamefont {J.~M.}\ \bibnamefont
  {Burgers}},\ }\href@noop {} {\bibfield  {journal} {\bibinfo  {journal} {Adv.
  in Appl. Mech.}\ }\textbf {\bibinfo {volume} {1}},\ \bibinfo {pages} {171}
  (\bibinfo {year} {1948})}\BibitemShut {NoStop}%
\bibitem [{\citenamefont {Rusanov}(1970)}]{rusanov1970method}%
  \BibitemOpen
  \bibfield  {author} {\bibinfo {author} {\bibfnamefont {V.}~\bibnamefont
  {Rusanov}},\ }\href {\doibase http://dx.doi.org/10.1016/0021-9991(70)90077-X}
  {\bibfield  {journal} {\bibinfo  {journal} {J. Comput. Phys.}\ }\textbf
  {\bibinfo {volume} {5}},\ \bibinfo {pages} {507 } (\bibinfo {year}
  {1970})}\BibitemShut {NoStop}%
\bibitem [{\citenamefont {Oppenheim}\ \emph {et~al.}(1983)\citenamefont
  {Oppenheim}, \citenamefont {Willsky},\ and\ \citenamefont
  {Nawab}}]{oppenheim1983signals}%
  \BibitemOpen
  \bibfield  {author} {\bibinfo {author} {\bibfnamefont {A.~V.}\ \bibnamefont
  {Oppenheim}}, \bibinfo {author} {\bibfnamefont {A.~S.}\ \bibnamefont
  {Willsky}}, \ and\ \bibinfo {author} {\bibfnamefont {S.~H.}\ \bibnamefont
  {Nawab}},\ }\href@noop {} {\emph {\bibinfo {title} {Signals and systems}}},\
  Vol.~\bibinfo {volume} {2}\ (\bibinfo  {publisher} {Prentice-Hall Englewood
  Cliffs, NJ},\ \bibinfo {year} {1983})\BibitemShut {NoStop}%
\bibitem [{\citenamefont {Pinton}\ \emph {et~al.}(2006)\citenamefont {Pinton},
  \citenamefont {Dahl},\ and\ \citenamefont {Trahey}}]{pinton2006rapid}%
  \BibitemOpen
  \bibfield  {author} {\bibinfo {author} {\bibfnamefont {G.~F.}\ \bibnamefont
  {Pinton}}, \bibinfo {author} {\bibfnamefont {J.~J.}\ \bibnamefont {Dahl}}, \
  and\ \bibinfo {author} {\bibfnamefont {G.~E.}\ \bibnamefont {Trahey}},\
  }\href@noop {} {\bibfield  {journal} {\bibinfo  {journal} {IEEE T. Ultrason.
  Ferr.}\ }\textbf {\bibinfo {volume} {53}},\ \bibinfo {pages} {1103} (\bibinfo
  {year} {2006})}\BibitemShut {NoStop}%
\end{thebibliography}%

\end{document}